\documentclass[preprint,aps,12pt,notitlepage,nofootinbib,tightenlines]{revtex4}
\usepackage{amsmath}
\usepackage{enumerate}
\usepackage{bm}
\usepackage{times}
\usepackage{braket}
\usepackage{color}
\usepackage{epsfig}
\usepackage{slashed}
\usepackage{hyperref}
\usepackage{multirow}
\usepackage{booktabs}
\usepackage{array}
\usepackage{float}
\newcommand{\X}{$X_{5/3}$ }
\newcommand{\beq}{\begin{eqnarray}}
\newcommand{\eeq}{\end{eqnarray}}
\newcommand{\be}{\begin{equation}\begin{aligned}}
\newcommand{\ee}{\end{aligned}\end{equation}}

\newcommand{\gev}{\text{GeV}}

\definecolor{Red}{rgb}{1.,0.,0.}

\definecolor{Blue}{rgb}{0.,0.,1.}

\definecolor{nicered}{rgb}{0.7,0.1,0.1}
\definecolor{nicegreen}{rgb}{0.1,0.5,0.1}
\def\lsim{ {\ \lower-1.2pt\vbox{\hbox{\rlap{$<$}\lower6pt\vbox{\hbox{$\sim$}}}}\ } }
\def\gsim{ {\ \lower-1.2pt\vbox{\hbox{\rlap{$>$}\lower6pt\vbox{\hbox{$\sim$}}}}\ } }

\hypersetup{colorlinks,citecolor=nicegreen,linkcolor=nicered}
\begin{document}
\title{Single production of vectorlike quarks with charge 5/3 at the 14 TeV LHC}
\author{Yao-Bei Liu\footnote{E-mail: liuyaobei@hist.edu.cn}, Bo Hu, Chao-Zheng Li}
\affiliation{Henan Institute of Science and Technology, Xinxiang 453003, P.R. China}
\begin{abstract}
 In a framework of the Standard
Model (SM) simply extended by an SU(2) doublet $\left(X,T\right)$  including a vectorlike
 $X$-quark~(VLQ-$X$), with electric charge $|Q_{X}| = 5/3$, we investigate the single production of the VLQ-$X$ induced by the couplings between the VLQ-$X$ with the first and the third generation quarks at the Large Hadron Collider (LHC) operating at $\sqrt{s}=14$~TeV. The signal is searched in events including same-sign dileptons (electrons or muons), one $b$-tagged jet and missing energy, where the $X$ quark is assumed to decay into a top quark and a W boson,  both decaying leptonically.
After a rapid simulation of signal and background events, the 95\% CL exclusion limits and the $5\sigma$ discovery reach  are respectively obtained at the LHC with an integrated luminosity
of 300 and 3000 fb$^{-1}$, respectively.
\end{abstract}

\maketitle

\newpage
\section{Introduction}
 To promote a potential solution to the gauge hierarchy problem~\cite{Susskind:1978ms}, new vectorlike quarks (VLQs)  with mass at the TeV scale are often present  in  many extensions of the Standard Model~(SM), such as little Higgs models~\cite{Arkani-Hamed:2002iiv,ArkaniHamed:2002qy,Han:2003wu,Chang:2003vs}, composite Higgs models~\cite{Agashe:2004rs,Contino:2006qr,Lodone:2008yy,Matsedonskyi:2012ym}, and other extended models~\cite{He:1999vp,Wang:2013jwa,He:2001fz,He:2014ora}, in which they cancel top-quark
loop contributions to the Higgs mass~\cite{Xiao:2014kba}. In the renormalizable
extensions of the SM, the canonical representation of VLQs generally constitutes one of seven
multiplets VLQs, including two singlet [$T$, $ B$], three doublets [ $\left(X,T\right),\left(T,B\right)$ or $\left(B,Y\right)$], and two triplets [$\left(X,T,B\right)$ or $\left(T,B,Y\right)$].  In the proposed model, $T$ and $B$ can be regarded  as the top and
bottom partners, respectively. $Y$ and
$X$ quarks have exotic electric charges of -4/3 and 5/3, respectively. Here we concentrate on the vectorlike $X$-quark~(VLQ-$X$), which commonly occurs as part of an $SU(2)_L \times SU(2)_R$ bi-doublet in models that preserve the custodial
symmetry~\cite{Agashe:2006at,Chivukula:2011jh}, and in models where some VLQs are added via renormalizable couplings~\cite{Buchkremer:2013bha,Aguilar-Saavedra:2013qpa}. Furthermore, such new particles could
generate characteristic signatures at the Large Hadron Collider~(LHC) and future high-energy colliders~(see, for example~\cite{Aguilar-Saavedra:2009xmz,Mrazek:2009yu,Dissertori:2010ug,Atre:2011ae,Cacciapaglia:2011fx,Cacciapaglia:2012dd,DeSimone:2012fs,Vignaroli:2012sf,Gopalakrishna:2013hua,Matsedonskyi:2014mna,Backovic:2014uma,
Chen:2016yfv,Fuks:2016ftf,Xie:2019gya,Aguilar-Saavedra:2019ghg,Belyaev:2021zgq,Bhardwaj:2022nko,Bhardwaj:2022wfz,Verma:2022nyd,Bardhan:2022sif,Han:2022jcp,Calabrese:2023ryr,Alves:2023ufm,Canbay:2023vmj,Belyaev:2023yym,Shang:2024wwy,Arhrib:2024tzm,Arhrib:2024dou,Yang:2024aav,Arhrib:2024nbj,Zhang:2024cpc}).

The ATLAS and CMS collaborations have conducted extensive searches for the pair production of VLQ-$X$~\cite{ATLAS:2018mpo,ATLAS:2018alq,ATLAS:2018ziw,CMS:2013wkd,CMS:2017mrm,CMS:2019eqb,Buckley:2020wzk}.  In the absence of any discovery, these
searches have put strong limits on VLQs masses according to the assumed decay pattern.
For instance, CMS Collaboration used an integrated luminosity 35.9~fb$^{-1}$ of data and provided the lower mass bounds about 1.30~(1.33)~TeV at 95\% confidence level (C.L.), for the case of left~(right)-handed couplings to $W$ bosons in a combination of the same-sign dileptons  and single-lepton final states~\cite{CMS:2018ubm}. Recently, the search was carried out on 139 fb$^{-1}$ of proton-proton collision data at $\sqrt{s}=13~$TeV with the ATLAS detector between
2015 and 2018 runs~\cite{ATLAS:2022tla}. This search excluded the presence of a VLQ-$X$  with mass up to 1.46 TeV for $\left(X,T\right)$ doublets for $Br(X \to tW)=1$.
 Besides, such VLQ-$X$ can also be singly produced at the LHC via  its electroweak~(EW) coupling, which is model-dependent and always depends on the EW coupling strength~\cite{Moretti:2016gkr,
Carvalho:2018jkq,Deandrea:2021vje}. Considering the final state including one muon or electron, the VLQ-$X$ with a
 relative width of 10\%, 20\%, and 30\% of its mass can be excluded below 0.92, 1.3, and 1.45 TeV, respectively at the $\sqrt{s}=13$ TeV LHC by the CMS Collaboration~\cite{CMS:2018dcw}.

 The upgrade of the
LHC to the high-luminosity phase (HL-LHC)~\cite{Apollinari:2015wtw} at center-of-mass
energy of 14 TeV with an integrated luminosity of 3000 fb$^{-1}$ will
extend the sensitivity and perspectives to  discover possible new physics signals.
  Due to the exotic charge, its main distinctive decay mode is $X\to t(\to W^{+}b)W^{+}$, which can give rise to the final states including same-sign two leptons~(SS2L) via the $W$ leptonic decays. In comparison with the existing searches for other channels, the SS2L channel has the great
advantage that most QCD backgrounds are gone, such as done in Refs.~\cite{Contino:2008hi,Ren:2017jbg,Zhou:2020ovl,Zhou:2020byj,Chiang:2021lsx,Yang:2021hcu,Cui:2022hjg,Arganda:2023pye}.
Considering VLQ-$X$ pair production and looking for a SS2L channel, the bounds on the VLQ-$X$ mass obtained from ATLAS~(CMS) are $M_{X}>670~(675)$ GeV
with a sample corresponding to an integrated luminosity of $4.7~(5)~\mathrm{fb}^{-1}$~\cite{ATLAS:2012hpa,CMS:2012mvl}.

\begin{figure}[htb]
\begin{center}
\vspace{-1.0cm}
\centerline{\epsfxsize=12cm \epsffile{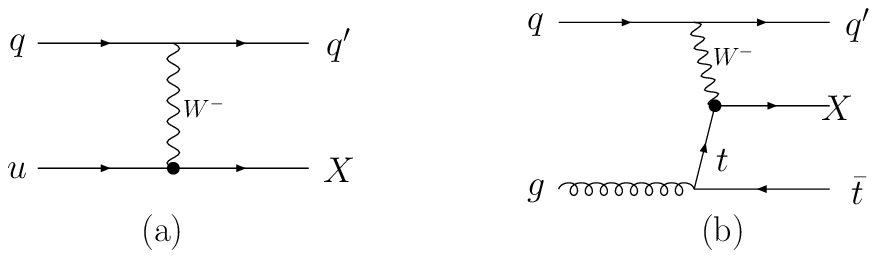}}
\vspace{-12.5cm}
\caption{Representative Feynman diagrams for the single production processes at the
LHC via couplings of the $X$ to (a) first generation quarks and (b) third generation quarks.}
\label{fey}
\end{center}
\end{figure}

Although the new VLQs are typically considered to mix sizably only with the third-generation of SM quarks, partial mixing to the first two generation SM quarks is not completely excluded~\cite{Buchkremer:2013bha}.
This is because cancellations among the effects of different types of new VLQs can
significantly alleviate the indirect constraints~\cite{delAguila:2000rc,Mehdiyev:2007pf,delAguila:2008iz,Atre:2008iu,Cacciapaglia:2010vn,Vignaroli:2012si,Bonne:2012im,Aguilar-Saavedra:2013wba,Ishiwata:2015cga}.
Especially, the EW precision bounds on the extra VLQs are generally weaker than the corresponding bounds on the extra chiral quarks such as the fourth family quarks/leptons~\cite{He:2001tp}. The crucial point is that even a small mixing to the first
generation may have a severe impact on single VLQs production processes  due to the presence of valence quarks in the initial state at the LHC~\cite{Beauceron:2014ila,Basso:2014apa,Liu:2016jho}. Very recently, the ATLAS Collaboration used an integrated luminosity 140~fb$^{-1}$ of data and excludes VLQs with masses below 1530~GeV for the branching ratio $Br(Q \to Wq)=1$\cite{ATLAS:2024zlo}.
In this work, we study the single production of the VLQ-$X$ at the HL-LHC in a simplified scenario where the VLQ-$X$ could mix with both the first-generation and the third-generation SM quarks,  as shown in  Fig.~\ref{fey}, and then analyze the SS2L final state via $X\to t(\to bW^{+})W^{+}$ decay channel followed by the $W$ leptonic decays mode.  Consequently, this
analysis complements the more common searches for VLQ-$X$ that are assumed to
mix only with the third-generation SM quarks~\cite{Shang:2023ebe,Han:2023jzm}.

The paper is arranged as follows. In Sec. II, we consider an effective model framework
including the VLQ-$X$ and calculate its single production cross sections at the 14 TeV LHC involving the mixing with both the first and third generate quarks. In Sec.~III, we discuss its observability via the decay mode $X\to  t(\to bW^{+})W^{+}\to \ell^{+}\ell^{+}b+\slashed E_T$
at the HL-LHC. Finally, conclusions are presented in Sec.~IV.

\section{VLQ-$X$ in a simplified model}
\subsection{An effective Lagrangian for doublet VLQ-$X$}
In general, the new fermions are assumed to interact with the SM fermions via Yukawa interactions, whose quantum numbers with respect to the weak $SU(2)_{L}\times U(1)_{Y}$ gauge group are thus limited by the requirement of an interaction with the Higgs doublet
and one of the SM fermions~(see for example \cite{Cacciapaglia:2011fx} and references therein). Here we focus on a specific simple model containing a vectorlike quark which is an $SU(2)$ doublet with the hypercharge $7/6$.  It is worth mentioning that such a new
doublet also contains a vectorlike $T$-quark with electric charge of $2/3$ due to its mixing with SM up-type quarks. The
Yukawa couplings induced by the new doublet $(X, U)^{T}$ generate a mixing in the up sector, with the three lighter mass
eigenstates identified with the SM quarks. Assuming that $u_{R}^{i}$ and $u_{L}^{i}$ represent the SM singlets and up
component of the doublets, the mass terms  are given by~\cite{Cacciapaglia:2012dd},
\beq
\mathcal{L}_{\rm mass} = - \sum_{i=1}^3\, \frac{y^i_u \upsilon}{\sqrt{2}}\, \bar u^i_L u^i_R -  \sum_{i=1}^3\, x_i\, \bar U_L u^i_R - M\, \bar U_L U_R  - M\, \bar X_L X_R  + h.c.\, ,
\eeq
where $y_{u}^{i}$ are the diagonal SM up Yukawa interactions, $\upsilon\sim246$~GeV is the Higgs vacuum expectation value (VEV), $x_i=\frac{\lambda^i \upsilon}{\sqrt{2}}$ represents the mixing generated by the Higgs VEV and $\lambda^i$ denote the new Yukawa couplings connecting the heavy quarks with the SM ones, $M$ is the vector-like quark mass.

We have not explicitly written down the heavy quark
Higgs couplings because they do not contribute appreciably to the production process of our
interest, and the details of the mixing matrices including all three families of quarks in the  SM  can be found in Refs.~\cite{Cacciapaglia:2011fx,Cacciapaglia:2012dd}.
As shown in Ref.~\cite{Cacciapaglia:2012dd}, the couplings of VLQ-$X$ with  the up-type SM quarks are generated by the coupling with its $SU(2)$ partner $U$, which mixes with the SM up
quarks. The $W$-boson interactions with the SM quarks and VLQ-$X$ are given by:
\beq
\mathcal{L}_{X} &=& i \frac{g}{\sqrt{2}}\, W^+_\mu\; \bar{X}_R \gamma^\mu \left( V_R^{43} t_R + V_R^{42} c_R + V_R^{41} u_R \right) + \nonumber \\
  & & i \frac{g}{\sqrt{2}}\, W^+_\mu\; \bar{X}_L \gamma^\mu \left( V_L^{43} t_L + V_L^{42} c_L + V_L^{41} u_L\right) + h.c.\,,
\eeq
Here we have neglected terms proportional to the vectorlike $T$-quark and the remaining terms can be found in~\cite{Cacciapaglia:2011fx}.
The matrix elements related to the new Yukawa couplings are given by
\beq
\begin{array}{cccc}
V_R^{41} = - \frac{x_1}{M}\,, & V_R^{42} = - \frac{x_2}{M}\,, & V_R^{43} = - \sin \theta_R\,; \\
V_L^{41} = - \frac{M_u x_1}{M^2}\,, & V_L^{42} = - \frac{M_c x_2}{M^2}\,, & V_L^{43} = - \frac{M_t}{M} \sin \theta_R\, .
\end{array}
\eeq
The above formulas show that the left-handed mixing with the light quarks, $V_L^{41}$ and $V_L^{42}$, can be safely neglected being
suppressed by the light quark masses~($M_{u}$ and $M_{c}$), while
the right-handed ones are proportional to the Yukawa masses $x_{i}$. Here we only simply recall the case for the VLQ-$X$ which is relevant for our discussion. Therefore, our study primarily concentrates on the right-handed coupling part of the interactions
involving the VLQ-$X$. Note that the matrices $V_L$ and $V_R$ will introduce a flavour mixing in the up sector, which can be strongly constrained by the flavor physics and the oblique parameters~\cite{Cao:2022mif,Alok:2015iha,Cacciapaglia:2015ixa,Vatsyayan:2020jan}.
 The stringent bound on flavour violation for this non-standard doublet case come from $D_0$--$\bar{D}_0$ mixing: $|V_R^{41}||V_R^{42}| < 3.2 \times 10^{-4}$~\cite{Cacciapaglia:2011fx}. However, the bounds on
the individual mixing terms, $V_R^{41}$ and $V_R^{42}$, are rather mild, i.e, $|V_R^{41}|<7.8 \times 10^{-2}$~\cite{Cacciapaglia:2011fx}, which is obtained by Atomic Parity Violation (APV)
experiments for the up,  and $|V_R^{42}| < 0.2$, which come from the measurement of $R_{c}$ at LEP~\cite{Cacciapaglia:2011fx}.

In our simplified model, we assume that the VLQ-$X$ couples only to the first- and third-generation
 quarks. An effective
Lagrangian framework for the interactions of a VLQ-$X$ with the SM quarks through the gauge boson $W$ exchange  is given by~\footnote{
Details are provided on the URL \href{http://feynrules.irmp.ucl.ac.be/wiki/VLQ\_xtdoubletvl}{feynrules.irmp.ucl.ac.be/wiki/VLQ\_xtdoubletvl}.}
\be
{\cal L}_{X} = g^{\ast}\left\{\sqrt{\frac{R_{L}}{1+R_{L}}} \frac{g}{\sqrt{2}} [\bar{X}_{R}W_{\mu}^{+}
    \gamma^{\mu} u_{R}]  + \sqrt{\frac{1}{1+R_{L}}}\frac{g}{\sqrt{2}} [\bar{X}_{R} W_{\mu}^{+} \gamma^{\mu} t_{R}] \right\}+ h.c.,
  \label{Xdoublet}
\ee
where
 $g$ is the $SU(2)_L$ gauge coupling constant,
 $g^{\ast}$ denotes the VLQ-$X$ coupling
strength to SM quarks in units of standard couplings, $R_L$ is the generation mixing coupling and two factors $\frac{R_{L}}{1+R_{L}}$ and $\frac{1}{1+R_{L}}$ describe the decay rate of the VLQ-$X$ to the first and third generation quark, respectively.  In the extreme case, $R_L\rightarrow 0$ and $R_L\rightarrow \infty$, respectively, correspond to
coupling to third-generation quarks and first-generation of quarks only.

Note that there are different symbols for the coupling coefficient in some literature~\cite{Shang:2023ebe,Han:2023jzm,Cao:2022mif}, such as $\sin \theta_{R}$, $\kappa$ and $g^{\ast}$. The relationship
 between these symbols of the coupling coefficient can be deduced as $\sin\theta_{R}=\kappa=g^{\ast}$ when $R_{L}=0$ in this work.
Assuming that the VLQs mix only with the third-generation quarks, the recent bound on the coupling parameter comes from the oblique parameters $S$, $T$ and $U$: the precise bound is mass dependent, i.e., $\sin\theta_{R}< 0.42~(0.25)$ for $M_{X}=1.3~(2.5)$~TeV in the $\left(X,T\right)$ doublet model~\cite{Cao:2022mif}.  The strong limit for the up quark mixing  come from the APV experiments: $|V_R^{41}|=g^{\ast}\sqrt{\frac{R_{L}}{1+R_{L}}}< 7.8 \times 10^{-2}$~\cite{Cacciapaglia:2011fx}. For a typical value $g^{\ast}=0.1~(0.2)$, one can get $R_{L}< 1.55~(0.18)$.
 Here we take only a phenomenologically guided limit and  choose a slightly looser range: $g^{\ast}\leq 0.5$ and  $0\leq R_L\leq 1$.

\subsection{Decay and production cross section}
\begin{figure}[htb]
\begin{center}
\vspace{-0.5cm}
\centerline{\epsfxsize=10cm \epsffile{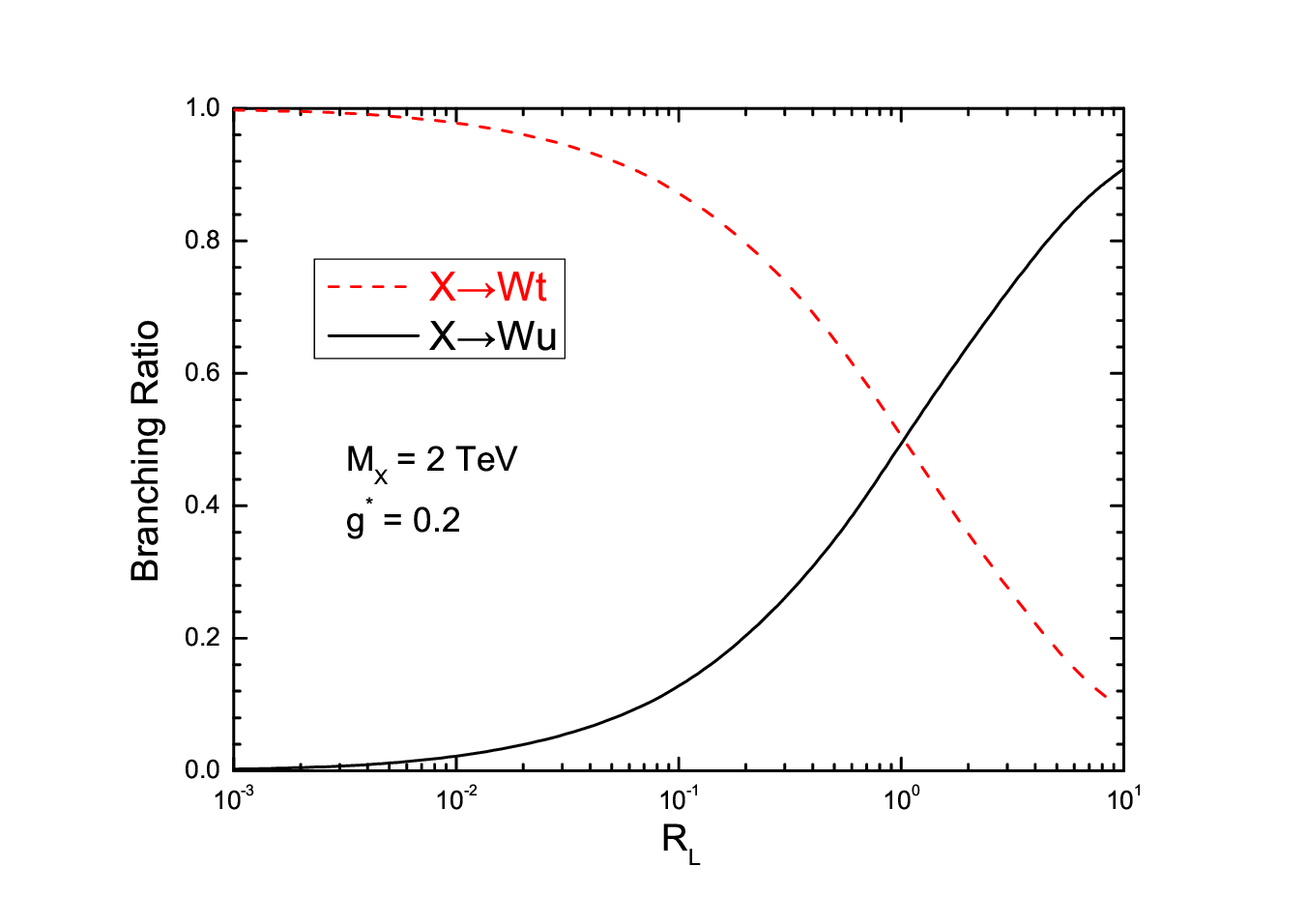}}
\caption{The branching ratio as a function of $R_L$ for $g^{\ast}=0.2$ and $M_{X}=2.0$~TeV. }
\label{fig2}
\end{center}
\end{figure}
\begin{figure}[htb]
\begin{center}
\vspace{-0.5cm}
\centerline{\epsfxsize=8cm \epsffile{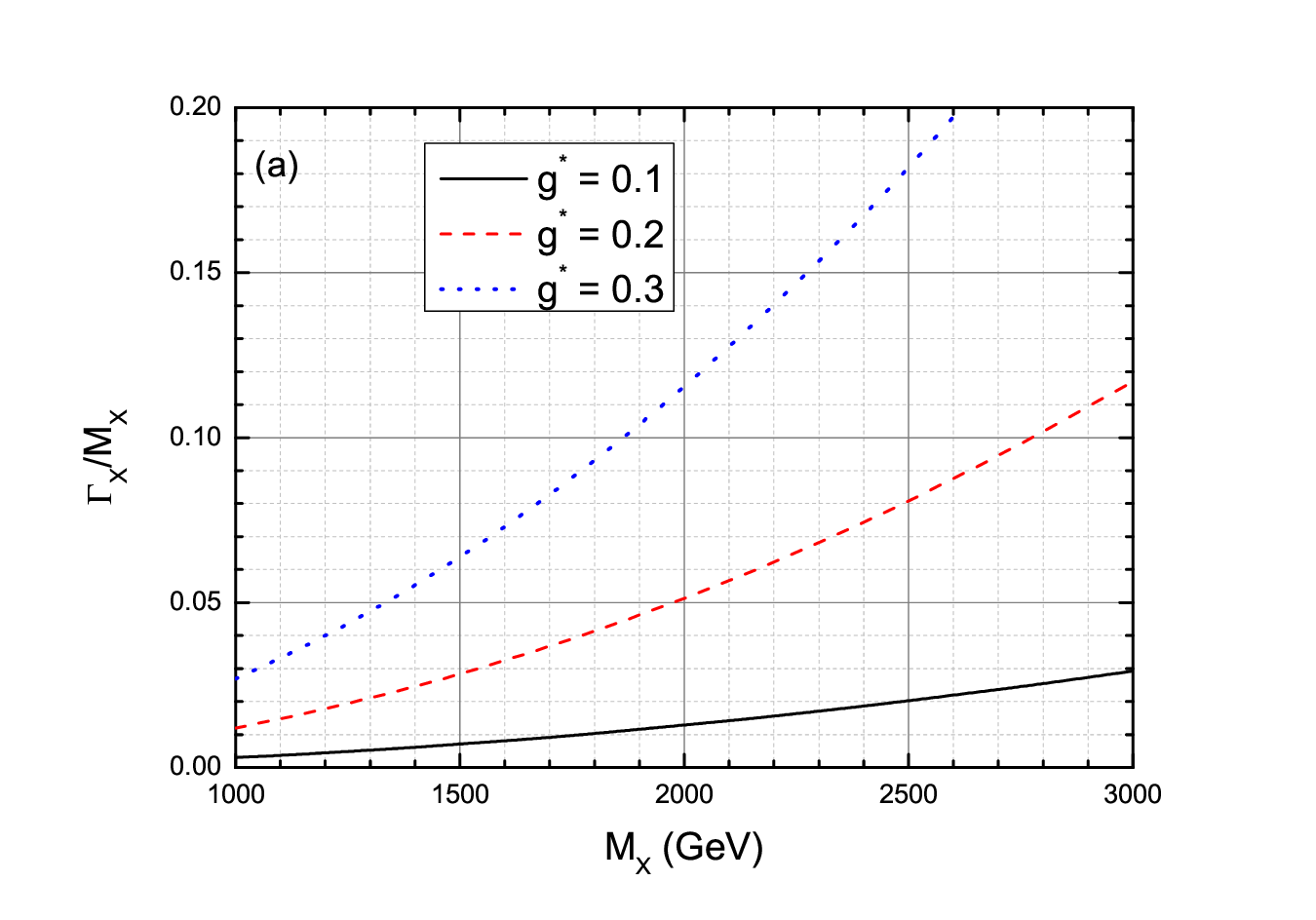}\hspace{-1.0cm}\epsfxsize=8cm \epsffile{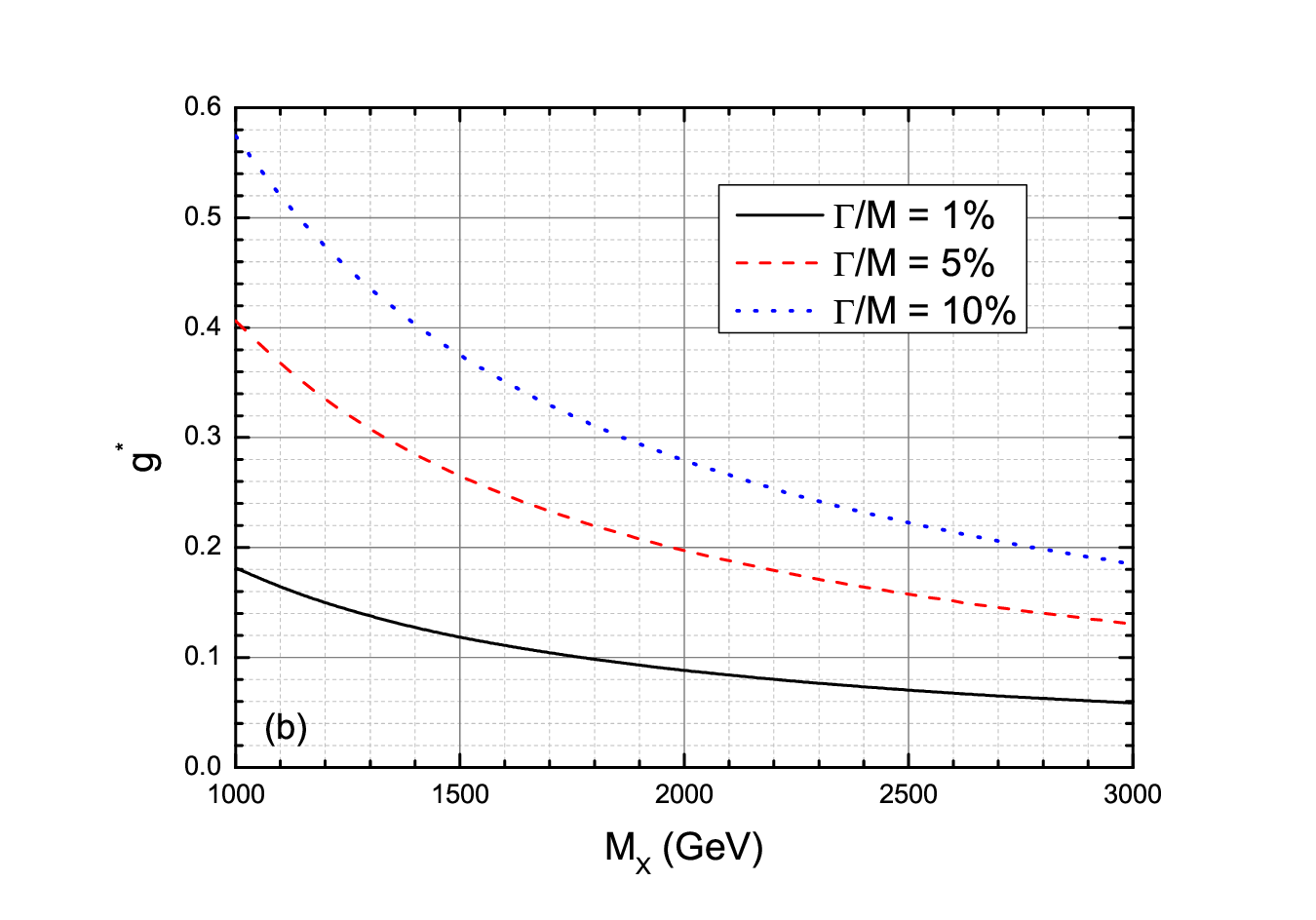}}
\caption{(a) The width-over-mass ratio $\Gamma_{X}/M_{X}$ as a function of $M_X$ for three typical coupling strength $g^{\ast}=0.1,0.2$ and 0.3. (b)The value of  $g^{\ast}$ as a function of $M_X$ for different width-over-mass ratio $\Gamma_{X}/M_{X}=1\%$, 5\% and 10\%.}
\label{fig3}
\end{center}
\end{figure}
Due to its charge, the VLQ-$X$ can decay only to $Wt$ and $Wu$ therefore
$BR(X\to Wt)+BR(X\to Wu) = 100\%$, as shown in Fig.~\ref{fig2}. Note that, when
$R_L$ up to 1.0, the decay rate of VLQ-$X$ to the first-generation quark reaches to the same value of VLQ-$X$ to
the third-generation quark decay.
 The decay width  depends on the coupling parameter $g^{\ast}$ and its mass $M_X$. For a fixed mass of VLQ-$X$, the total width $\Gamma_{X}$ is always proportional to $(g^{\ast})^{2}$, as shown in Fig.~\ref{fig3}(a). Therefore, the coupling strength $g^{\ast}$ can also be fixed to obtain  a specific width-over-mass ratio $\Gamma_{X}/M_X$, as shown in Fig.~\ref{fig3}(b).  One can see that, for the scenario $\Gamma_{X}/M_X=10\%$, $g^{\ast}$ is approximately smaller than 0.2 in the whole range of explored masses.
 It has been pointed out in~\cite{Barducci:2013zaa} that the Breit-Wigner (BW) form of a propagator
may be appropriate for narrow resonances where the width-over-mass ratio is smaller than
10\%. Thus the VLQ-$X$ can be assumed with narrow decay widths  and the production and decay can be factorised under the narrow-width approximation (NWA) case~\footnote{See Refs.~\cite{Moretti:2016gkr,
Carvalho:2018jkq,Deandrea:2021vje} for large-width effects in vector-like quark production.}.

\begin{figure}[thb]
\begin{center}
\vspace{-0.5cm}
\centerline{\epsfxsize=8cm \epsffile{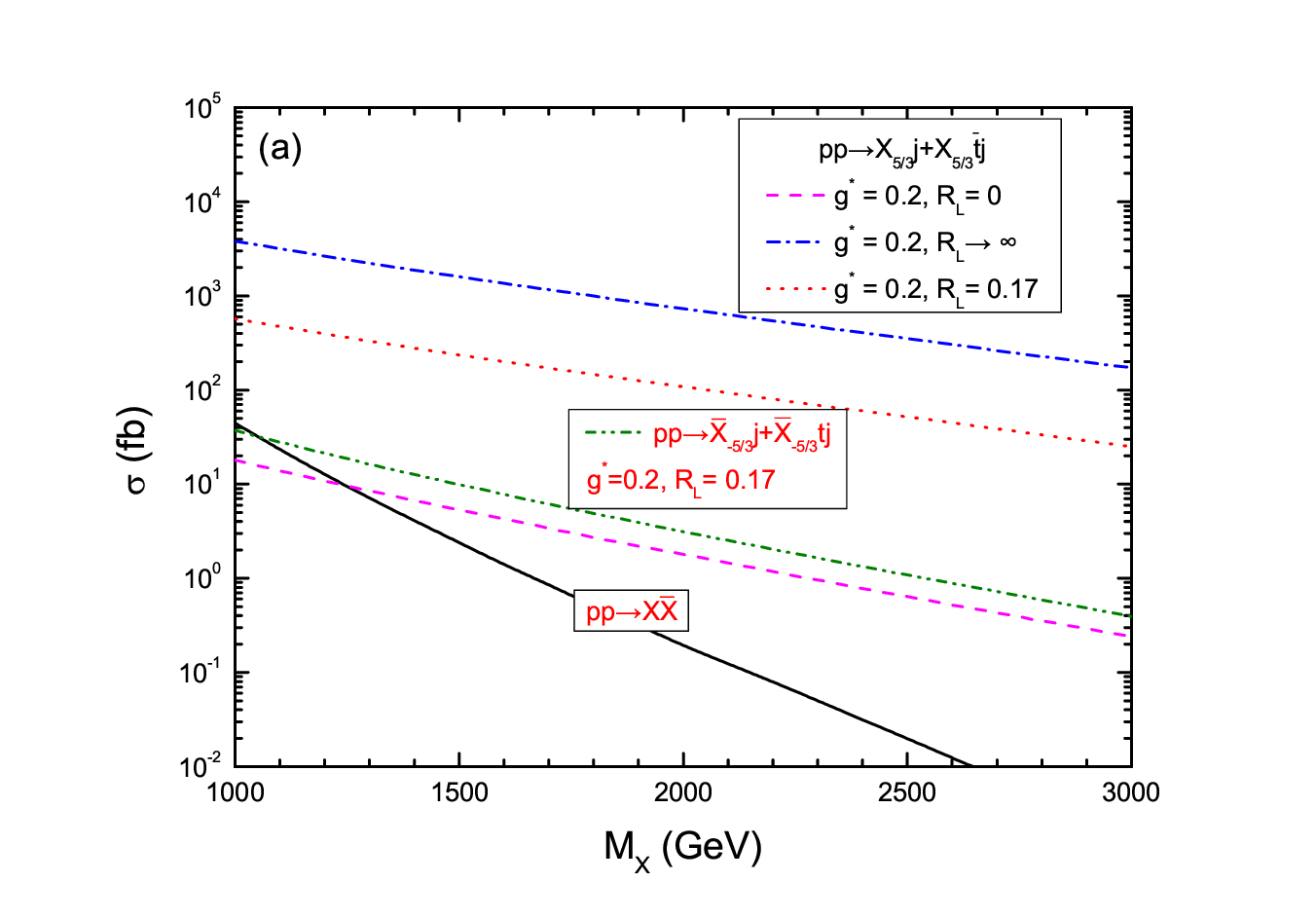}\hspace{-1.0cm}\epsfxsize=8cm \epsffile{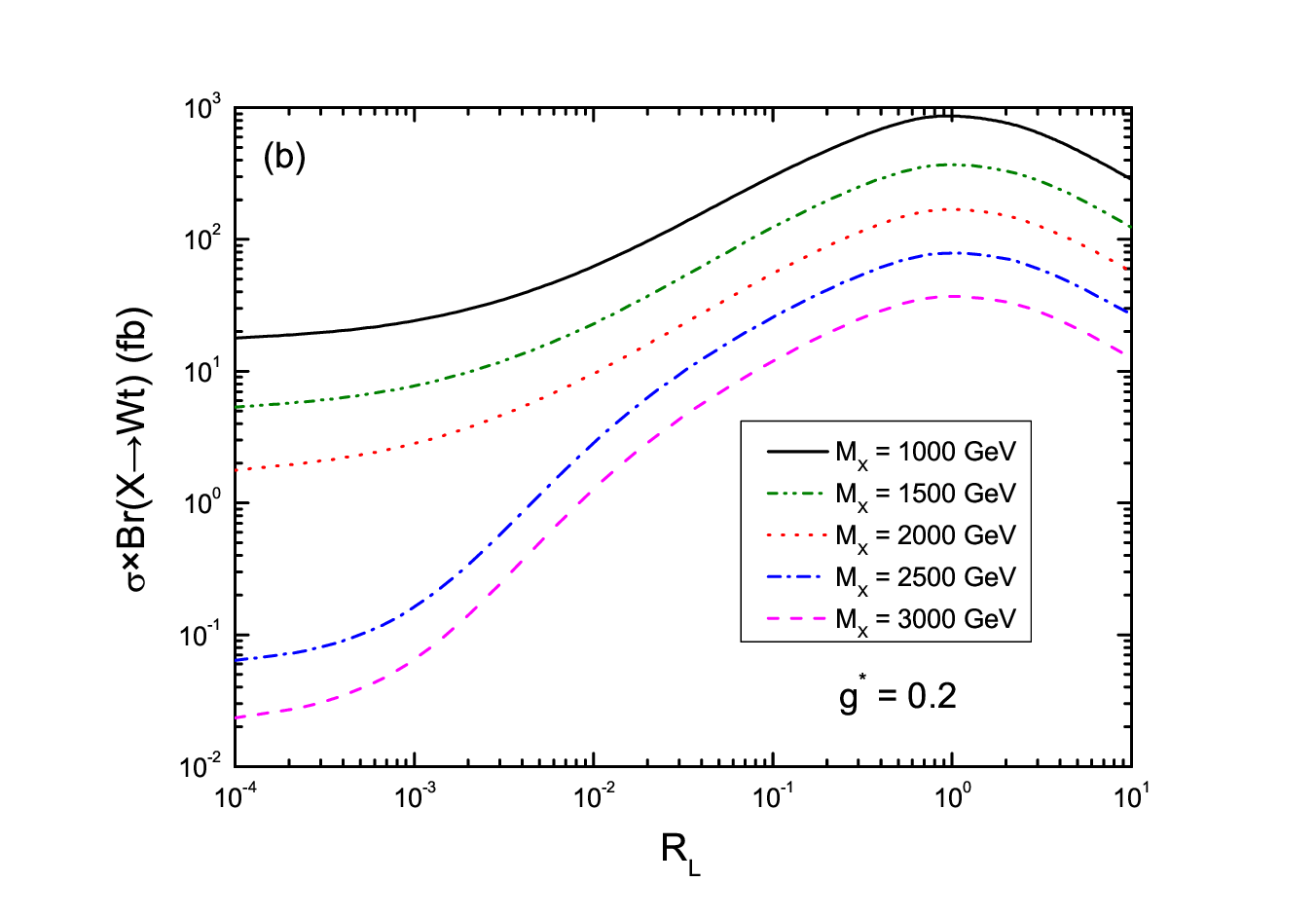}}
\caption{(a) Cross sections of VLQ-$X$ as a function of the mass $M_{X}$ for single and pair production processes at the 14~TeV LHC.
(b) Production cross section of $\sigma\times Br(X\to tW)$ as a function of $R_{L}$ for $g^{*}=0.2$ and five typical values of $M_{X}$ at the 14~TeV LHC.}
\label{cross}
\end{center}
\end{figure}
The VLQ-$X$ can be both single and pair produced at the LHC.
The cross sections of single VLQ-$X$ production versus
its mass at the 14 TeV LHC has been presented in Fig.~\ref{cross}(a) with different coupling parameters. The leading-order~(LO) cross sections are obtained using MadGraph5-aMC$@$NLO~\cite{Alwall:2014hca} with default NN23L01 parton distribution function (PDFs)~\cite{Ball:2014uwa} taking the default renormalization and factorization scales.   One can note that pair production
 is only dominant at low masses of the VLQ-$X$ but decreases faster than single production when the
mass of the VLQ-$X$ increases. This is due to the phase space suppression and the decrease of the parton distribution functions with the centre of mass energy of the
parton-level collision. However, single
heavy quark production has the advantage of less phase-space suppression and longitudinal
gauge boson enhancement of order $M_{X}^{2}/M_{W}^{2}$
at higher energies compared to pair production.
With the fixed parameters $g^{*}=0.2$ and  $R_{L}=0.17$, the cross section can reach about 236~(108) fb  for $M_X=1.5~(2.0)$ TeV.
We also plot the cross section for the process $pp\to \bar{X}j+\bar{X}tj$ for the same fixed coupling parameters. One can see that the production cross section of $pp\to Xj+X\bar{t}j$ is much larger than that for the conjugate process due to the difference in the PDFs of valence and sea quarks in the initial states.

In Fig.~\ref{cross}(b), we also show the dependence of the cross sections  $\sigma\times Br(X\to tW)$  on the mixing parameter $R_L$.
We generate  five benchmark points varying the VLQ-$X$ mass in steps of 500 GeV in the range [1000;~3000] GeV with $g^{\ast}=0.2$.
One can see that (i) in the range of $R_L<1$, the production cross section increases largely with the increase of $R_L$. (ii) For $R_L>1$, the production cross section  will become small with the increase of $R_L$. This effect is mainly due to the increased admixture of valence quarks in production, mitigated by a reduced $X\to tW$ branching ratio with increasing $R_L$. The
cross section will reach a maximum for $R_L\simeq 1$, which corresponds
to $50\%-50\%$ mixing.
\section{Event generation and discovery potentiality}

In this section, we analyze the 14 TeV LHC observation potential by performing a
Monte Carlo  simulation of the signal and SM background events and explore
 the sensitivity to the VLQ-$X$  through the process
 \be
 pp\to &X(\to tW^{+})j \to t(\to bW^{+}\to b\ell^{+} \nu_{\ell})W^{+}(\to \ell^{+} \nu_{\ell}) j,\\
  pp\to &X(\to tW^{+})\bar{t}j \to t(\to bW^{+}\to b\ell^{+} \nu_{\ell})W^{+}(\to \ell^{+} \nu_{\ell})\bar{t} j,
\ee
where $\ell= e, \mu$.

In this analysis, the final states with SSL~(muon or electron), one $b$-tagged jet, and missing transverse energy $\slashed E_T$
are studied. Note that we do not consider the reconstruction or selection for the associated anti-top quark
as well as
the leptons or jets originating from its decay, due to their much lower transverse momenta $p_T$. The dominant SM backgrounds come from the SM processes
  $t\bar{t}W^{+}$, $W^{+}W^{+}+$jets  and the nonprompt leptons (mainly from events with jets of heavy flavor, such as $t\bar{t}$).
To account for contributions from higher-order QCD corrections,
 the cross sections of dominant backgrounds at LO are adjusted to NLO or next-NLO (NNLO) order using
$K$-factors,  which are listed in table~\ref{kf}.

\begin{table}[htb]
\centering %
\caption{$K$-factors of the leading SM background processes for our analysis. \label{kf}}
\vspace{0.8cm}
\begin{tabular}{p{1.8cm}<{\centering} p{2.4cm}<{\centering} p{2.0cm}<{\centering} p{2.0cm}<{\centering} }
\toprule[1.5pt]
\hline
Process& $W^{+}W^{+}jj$&$t\bar{t}W^{+}$&$t\bar{t}$ \\  \hline
$K$-factor&1.04~\cite{Jager:2009xx,Melia:2010bm}&1.22~\cite{Campbell:2012dh}&1.6~\cite{Czakon:2013goa}\\
\hline
\end{tabular}
\end{table}

Signal and background events are generated at LO  using
MadGraph5-aMC$@$NLO. To perform the parton
shower and fast detector
simulations,  we transmit the parton-level events to Pythia8 \cite{pythia8} for parton showering and hadronization,  then processed through  Delphes 3.4.2~\cite{deFavereau:2013fsa} for detector simulation by using the standard HL-LHC detector parameterization shipped with the program.
 Finally, the resulting signal and background events are analyzed using MadAnalysis5~\cite{ma5}.

To identify objects, the following basic cuts are chosen at parton level:
 \be
p_{T}^{\ell/j}>~30~\gev,\quad
 |\eta_{\ell}|<~2.5, \quad
  |\eta_{j}|<~5, \quad
 \Delta R_{ij} > 0.4,\\
  \ee
where $\Delta R=\sqrt{\Delta\Phi^{2}+\Delta\eta^{2}}$ is the separation in the rapidity-azimuth plane and $p_{T}^{\ell/ j}$ and $|\eta_{\ell/j}|$ are the transverse momentum and pseudorapidity of the leptons and jets, respectively.
\begin{figure*}[htb]
\begin{center}
\centerline{\hspace{2.0cm}\epsfxsize=9cm\epsffile{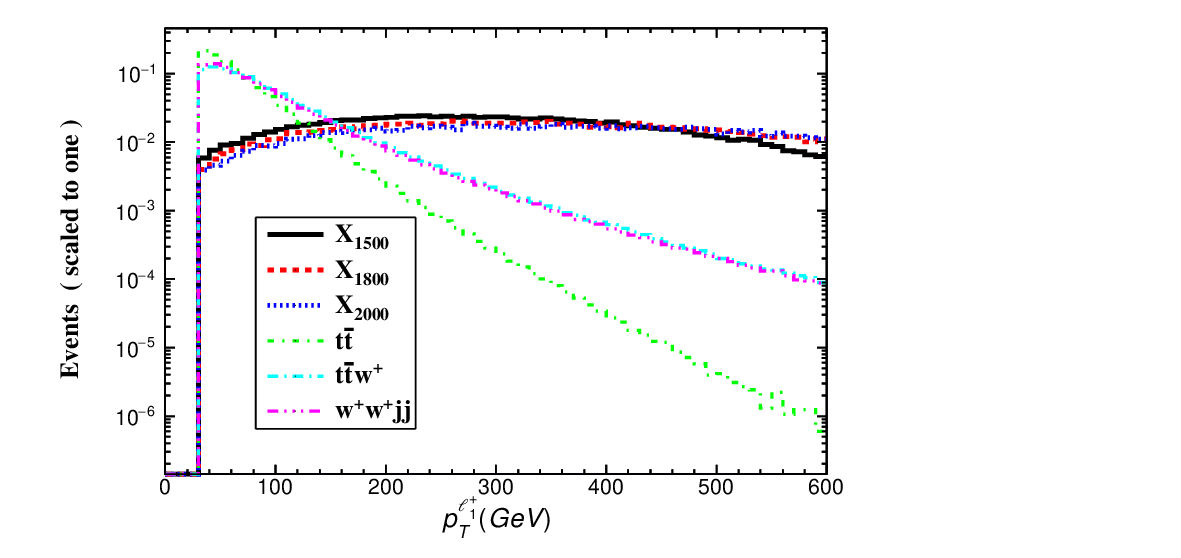}
\hspace{-2.0cm}\epsfxsize=9cm\epsffile{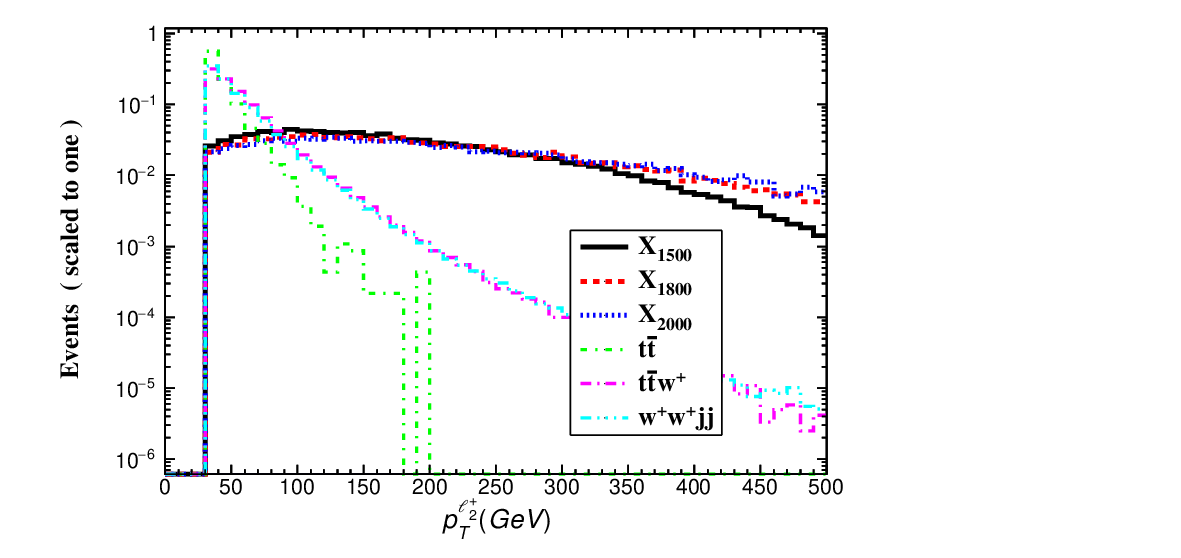}}
\centerline{\hspace{2.0cm}\epsfxsize=9cm\epsffile{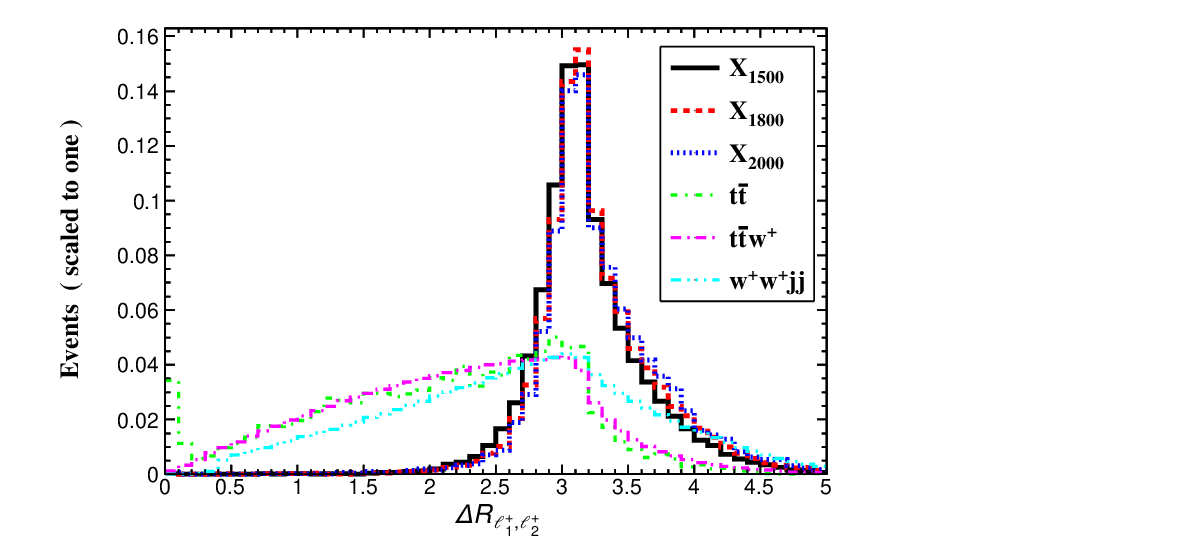}
\hspace{-2.0cm}\epsfxsize=9cm\epsffile{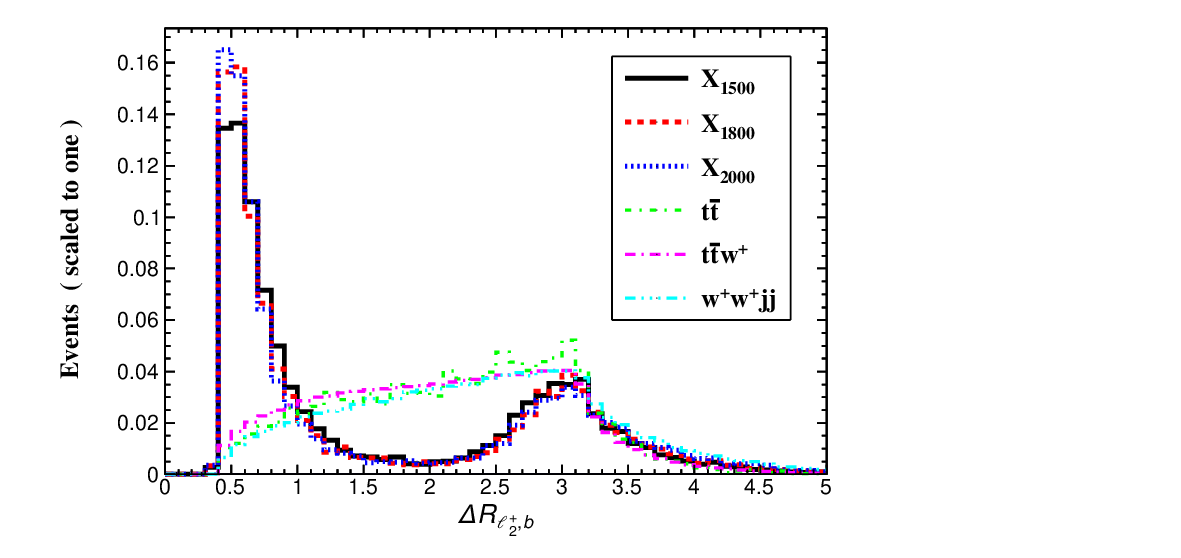}}
\centerline{\hspace{2.0cm}\epsfxsize=9cm\epsffile{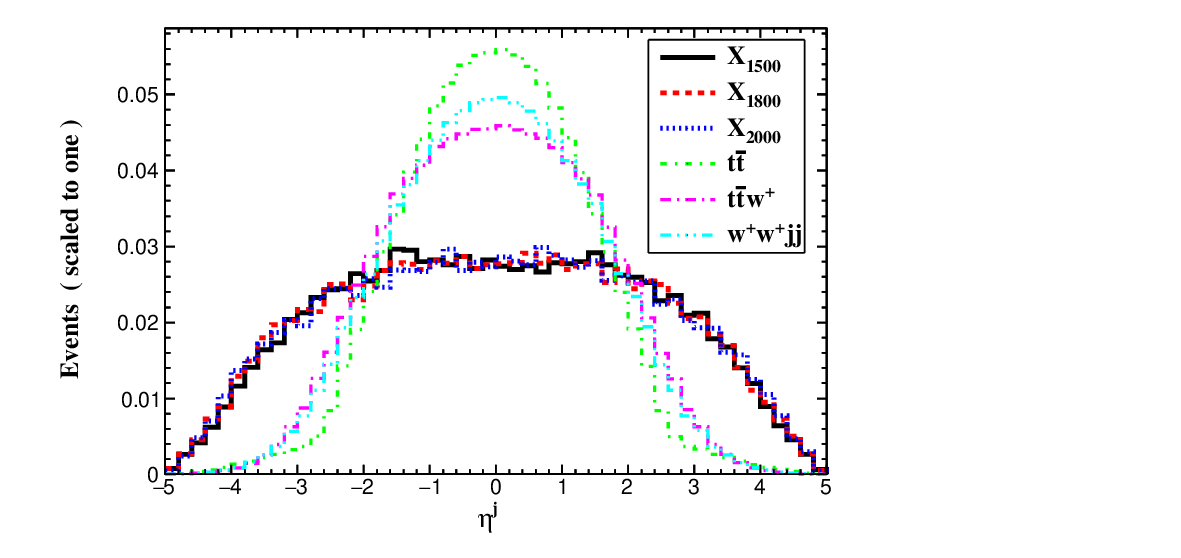}
\hspace{-2.0cm}\epsfxsize=9cm\epsffile{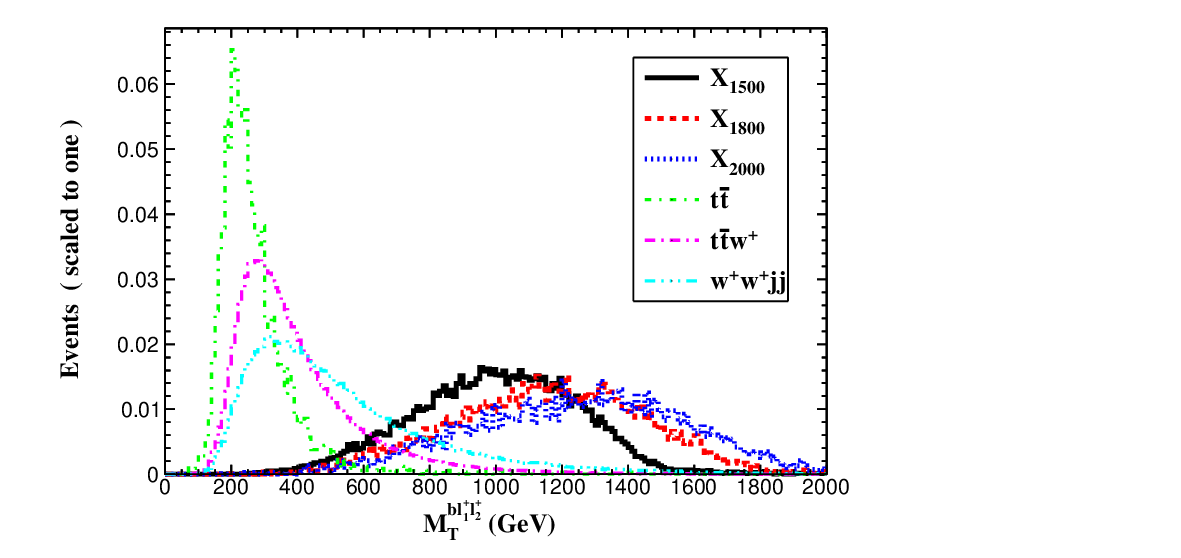}}
\caption{Normalized distributions for the signals (with $M_{X}=1500, 1800$ and 2000 GeV) and SM backgrounds. }
\label{fig5}
\end{center}
\end{figure*}

In Fig.~\ref{fig5}, we plot some differential distributions  for  signals and SM backgrounds at the LHC, such as the transverse momentum distributions of the leading and subleading leptons~($p_{T}^{\ell_{1,2}}$),  the separations $\Delta R_{\ell_{1},\ell_{2}}$ and $\Delta R_{\ell_{2},b}$, the rapidity of the forward jet and the transverse mass~\footnote{The definition of the transverse mass of the system can be seen from Ref.~\cite{Conte:2014zja}.} distribution for VLQ-$X$ $M_{T}^{b\ell_{1}\ell_{2}}$. Due to the larger mass of VLQ-$X$, the decay products are highly boosted, and thus the $p_{T}^{\ell}$ peaks of the signals are larger than those of the SM backgrounds.
Based on these kinematical distributions, we apply
the following kinematic cuts to the events to distinguish the signal from the SM backgrounds.
 \begin{itemize}
\item Cut 1: Exactly two same-sign isolated leptons~$[N(\ell^{+})=2]$  and the transverse momenta of the leading and subleading leptons are required $p_{T}^{\ell_{1,2}}> 120~(60) \rm ~GeV$, the distance between two leptons lies in $\Delta R_{\ell_{1},\ell_{2}}> 2.5$.
\item Cut 2: At least one $b$-tagged jet~$[N(b)\geq 1]$ with $p_{T}^{b}> 30 \rm ~GeV$. Besides, the distance between the subleading leptons with the $b$-tagged jet is required $\Delta R_{\ell_{2},b}< 1$.
\item Cut 3: Since the jet from splitting of a valence quark with one $W$ emission always has a strong
forward nature, the light untagged jet is required to have $\mid\eta_{j}\mid > 2$.
\item Cut 4: The transverse mass of final system is required to have $M_{T}^{b\ell_{1}\ell_{2}}> 600 \rm ~GeV$.
\end{itemize}

\begin{table}[htb]
\centering %
\caption{Cut flow of the cross sections (in fb) for the signals and SM backgrounds at the 14 TeV LHC and three typical VLQ-$X$ masses. Here we set a benchmark value of $g^{*}=0.2$ and $R_{L}=0.17$. \label{cutflow}}
\vspace{0.8cm}
\begin{tabular}{p{1.6cm}<{\centering} p{2.0cm}<{\centering} p{2.0cm}<{\centering} p{2.0cm}<{\centering}p{0.3cm}<{\centering} p{2.2cm}<{\centering}  p{2.2cm}<{\centering} p{2.2cm}<{\centering} }
\toprule[1.5pt]
 \multirow{2}{*}{Cuts}& \multicolumn{3}{c}{Signals}&\multicolumn{4}{c}{Backgrounds}  \\ \cline{2-4}  \cline{6-8}
&1500~GeV&1800 GeV &2000 GeV&& $t\bar{t}$  &$t\bar{t}W^{+}$  &$W^{+}W^{+}jj$\\    \cline{1-8} \midrule[1pt]
Basic&6.18&3.88&2.86&&18741&11.1&2.73\\
Cut 1&2.44&1.38&0.94&&0.1&0.32&0.11\\
Cut 2&0.78&0.44&0.28&&0.01&0.06&$8\times 10^{-4}$\\
Cut 3 &0.37&0.20&0.13&&0.0022&0.009&$1\times 10^{-4}$\\
Cut 4 &0.36&0.197&0.125&&0.0011&0.0046&$7.7\times 10^{-5}$\\
\hline
\end{tabular}
 \end{table}

In Table~\ref{cutflow}, we present the cross sections of three typical signal ($M_X=1500, 1800, 2000$~GeV) and the relevant
backgrounds after imposing
the above mentioned cuts.  Notably,  all background processes are suppressed very significantly at the end of the cut flow, and the dominant SM  background comes from the $t\bar{t}W^{+}$ process, with a cross section of $4.6\times 10^{-3}$~fb.

The following statistical significance is used  to estimate the expected discovery and exclusion limits~\cite{Cowan:2010js}
\be
\mathcal{Z}_\text{disc} &=
  \sqrt{2\left[(s+b)\ln\left(\frac{(s+b)(1+\delta^2 b)}{b+\delta^2 b(s+b)}\right) -
  \frac{1}{\delta^2 }\ln\left(1+\delta^2\frac{s}{1+\delta^2 b}\right)\right]} \\
   \mathcal{Z}_\text{excl} &=\sqrt{2\left[s-b\ln\left(\frac{b+s+x}{2b}\right)
  - \frac{1}{\delta^2 }\ln\left(\frac{b-s+x}{2b}\right)\right] -
  \left(b+s-x\right)\left(1+\frac{1}{\delta^2 b}\right)},
 \ee
with
 \be
 x=\sqrt{(s+b)^2- 4 \delta^2 s b^2/(1+\delta^2 b)},
 \ee
 where  $s$ and $b$ denote the event number of signal and background  after the above cuts, respectively.  The integrated luminosity
at the 14 TeV LHC is set at 300 and 3000~fb$^{-1}$, respectively.   $\delta$ denotes the percentage systematic uncertainty that inevitably appears in the measurement of the SM background.
In the limit case ($\delta \to 0$),  these expressions  can be simplified as
\be
\mathcal{Z}_\text{disc} &= \sqrt{2[(s+b)\ln(1+s/b)-s]},\\
 \mathcal{Z}_\text{excl} &= \sqrt{2[s-b\ln(1+s/b)]},
 \ee
 as already used in many of the phenomenological studies.

\begin{figure}[ht]
\begin{center}
\vspace{1.5cm}
\centerline{\epsfxsize=9cm \epsffile{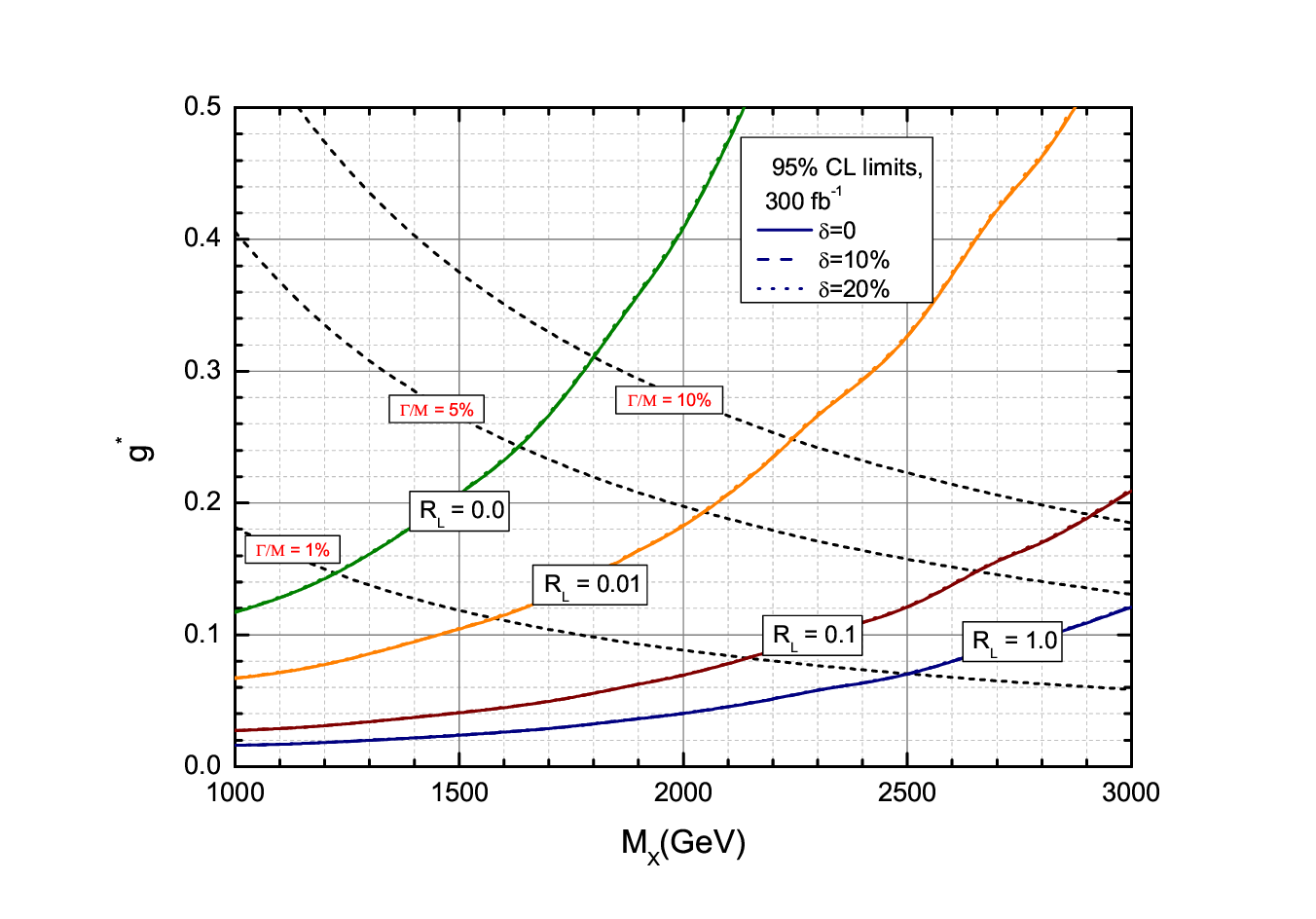}\epsfxsize=9cm \epsffile{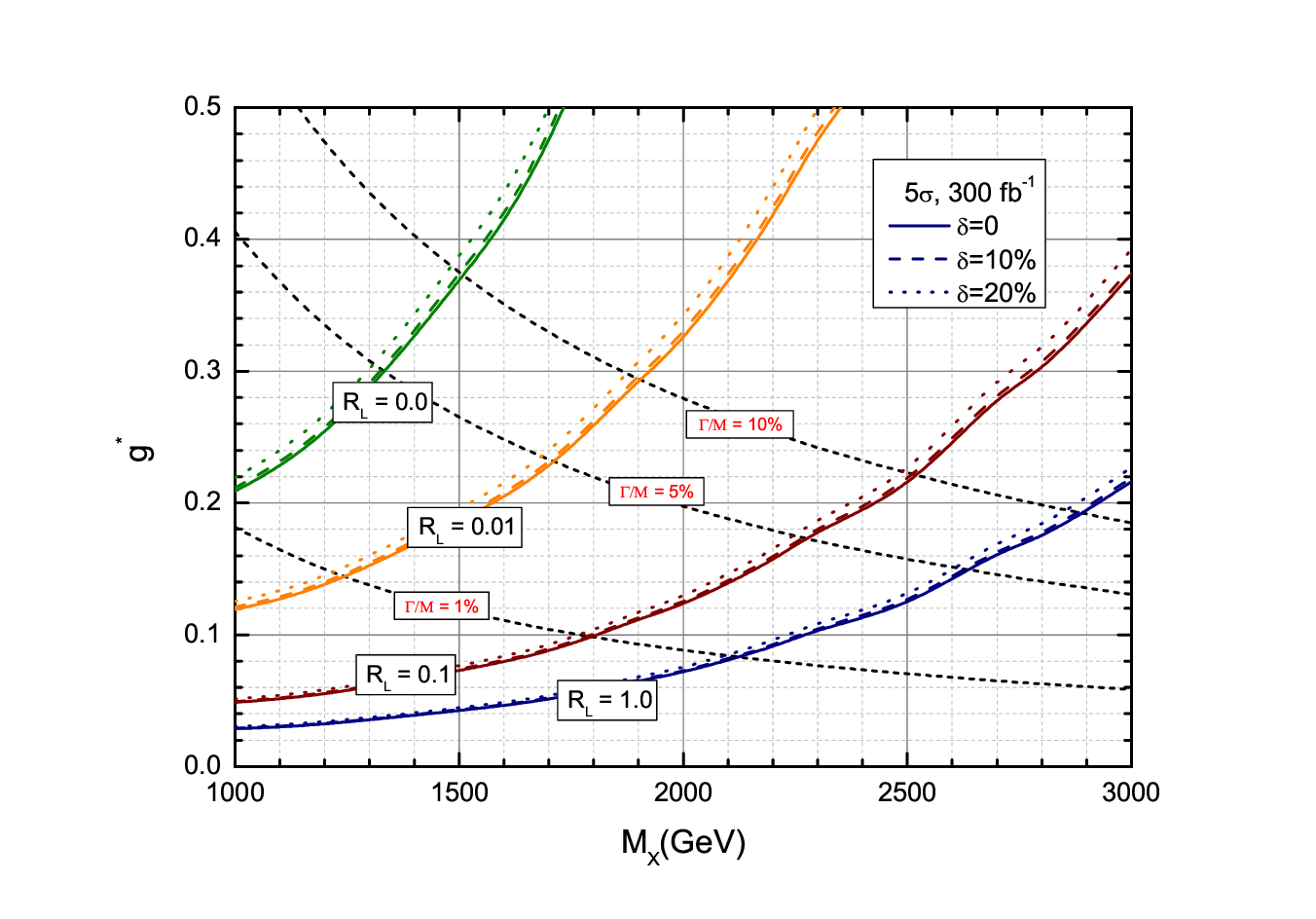}}
\centerline{\epsfxsize=9cm \epsffile{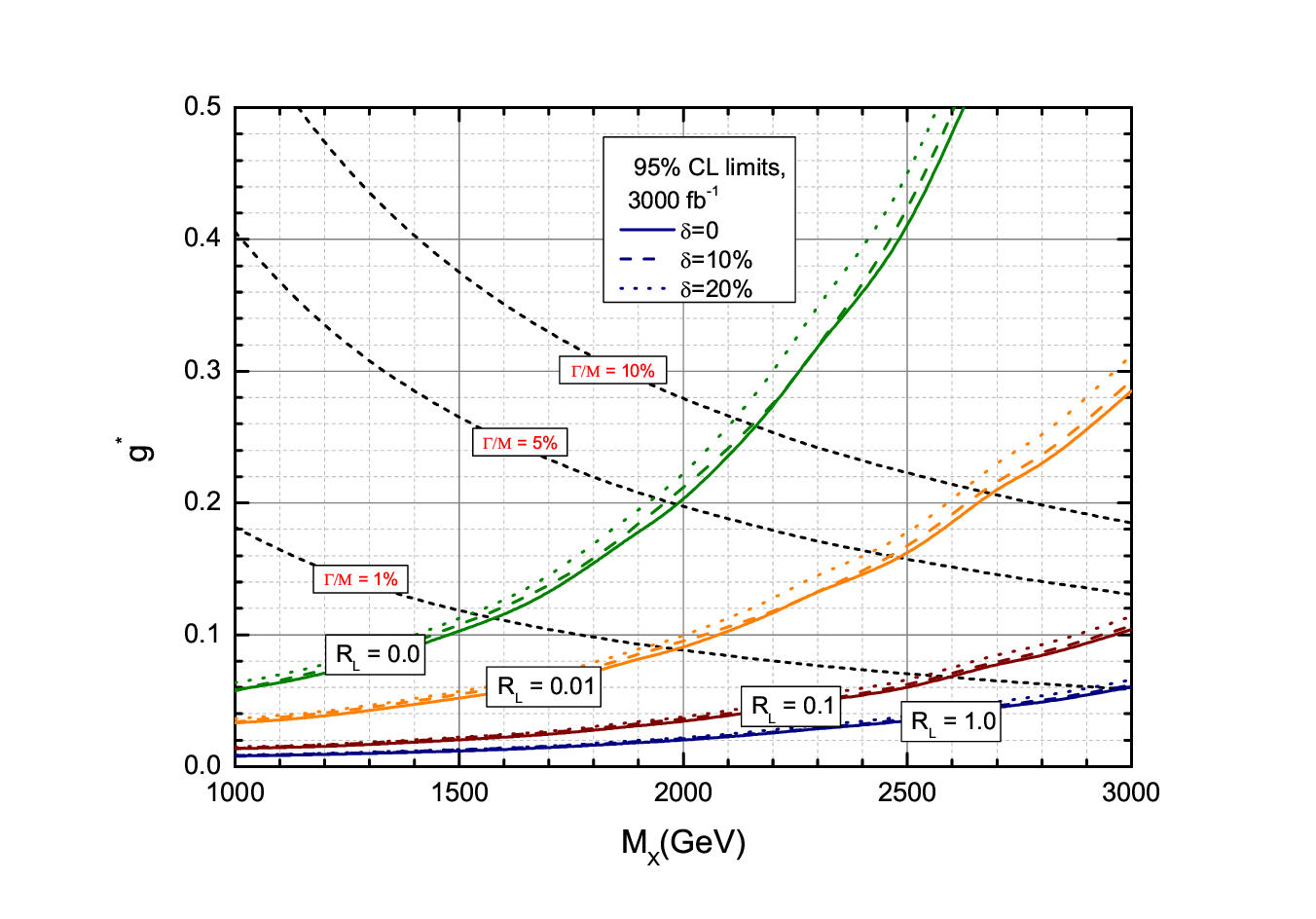}\epsfxsize=9cm \epsffile{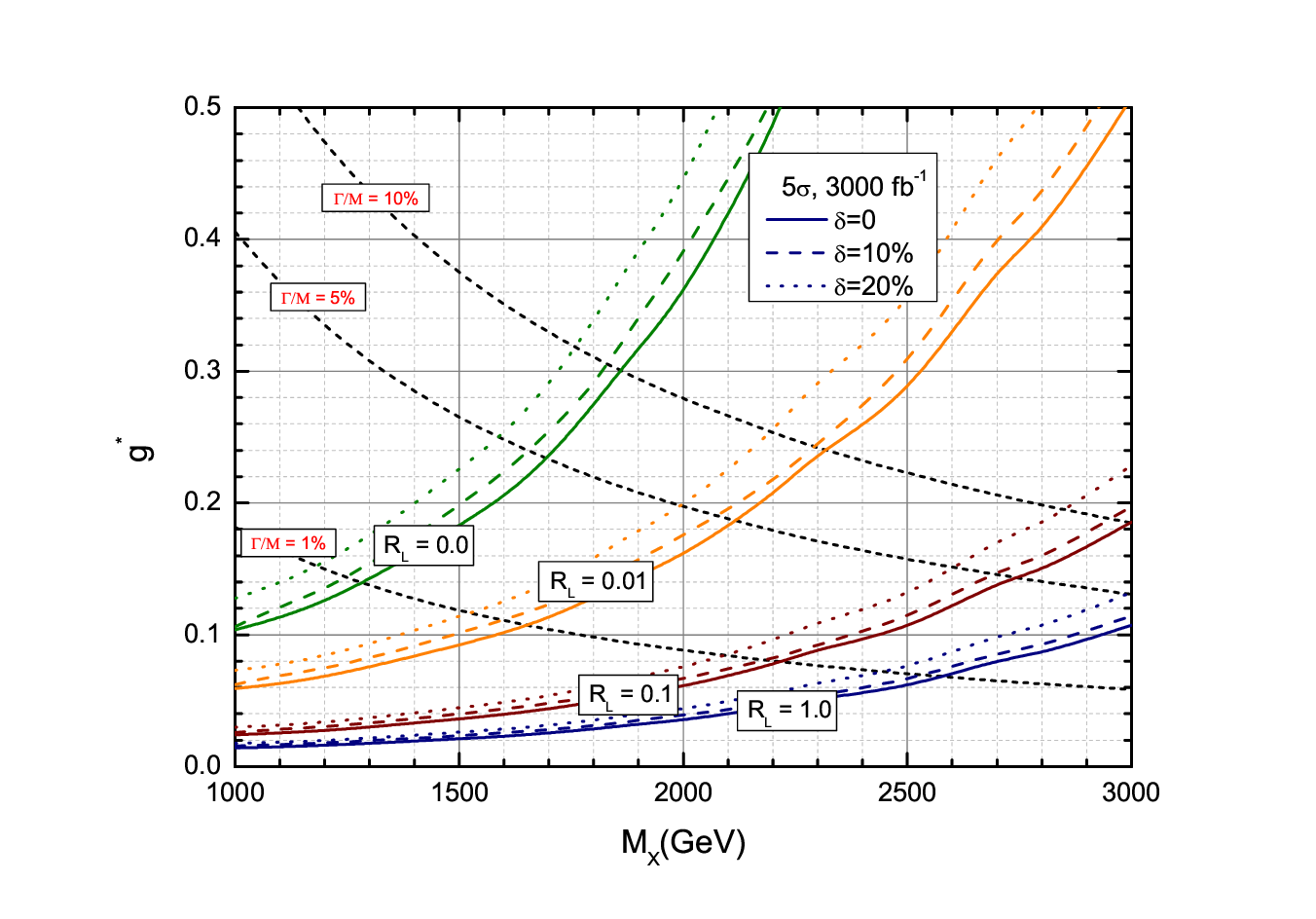}}
\caption{95\% CL exclusion limit (left panel) and $5\sigma$ (right panel) contour plots for the signal in $g^{*}-M_X$ at the 14 TeV LHC with an integrated luminosity of 3000 fb$^{-1}$ (upper) and 300 fb$^{-1}$ (down). Short-dashed lines denote the contours of $\Gamma_{X}/M_{X}$.}
\label{ss-mass}
\end{center}
\end{figure}

In Fig.~\ref{ss-mass}, we plot the excluded the 95\% CL exclusion limit and $5\sigma$ discovery reaches in the plane of $g^{*}-M_X$ at the 14 TeV LHC with an integrated luminosity of 300 fb$^{-1}$ and 3000 fb$^{-1}$ respectively for three different systematic
uncertainty cases: no systematics ($\delta=0$) , a mild systematic of $\delta=10\%$, and  a possible systematic of $\delta=20\%$. One can see that with a possible uncertainty of
20\%, sensitivities are slightly weaker than those with a mild systematic uncertainty of 10\% and no systematics of $\delta=0$.
In the presence of 10\% systematic uncertainty and $R_{L}=0$, the discovered (with $5\sigma$ level) regions are  $g^{*}\in [0.21, 0.4]~([0.1,0.4]))$ and $M_{X}\in [1000, 1550 ]~([1000,2000])$~GeV at the 14 TeV LHC with an integrated luminosity
of 300~(3000) fb$^{-1}$. Out of a discovery, the VLQ-$X$ can be excluded (at 95\% CL limits) in the  correlated
parameter space of $g^{*}\in [0.12, 0.4]~([0.06,0.4])$ and $M_{X}\in [1000, 1980 ]~([1000, 2450 ])$~GeV
for the same integrated luminosity. Assuming the non-vanishing $R_L$ value, i.e., $R_L=0.1$,  the discovery region can reach $g^{*}\in [0.05, 0.39]~([0.025,0.2]))$ and $M_{X}\in [1000, 3000 ]$~GeV with an integrated luminosity
of 300~(3000) fb$^{-1}$. Otherwise, the 95\% CL excluded region for the coupling parameter  is
$g^{*}\in [0.03, 0.21]~([0.015,0.11])$ and $M_{X}\in [1000, 3000 ]$~GeV with the same integrated luminosity at the 14 TeV LHC. Besides, although the vector
like quark width plays a significant role in their single
production, the region in this study is almost located
in $\Gamma_{X}/M_{X}<10\%$, and thus the NWA is reasonable
in our study.

\begin{figure}[ht]
\begin{center}
\vspace{1.5cm}
\centerline{\epsfxsize=9cm \epsffile{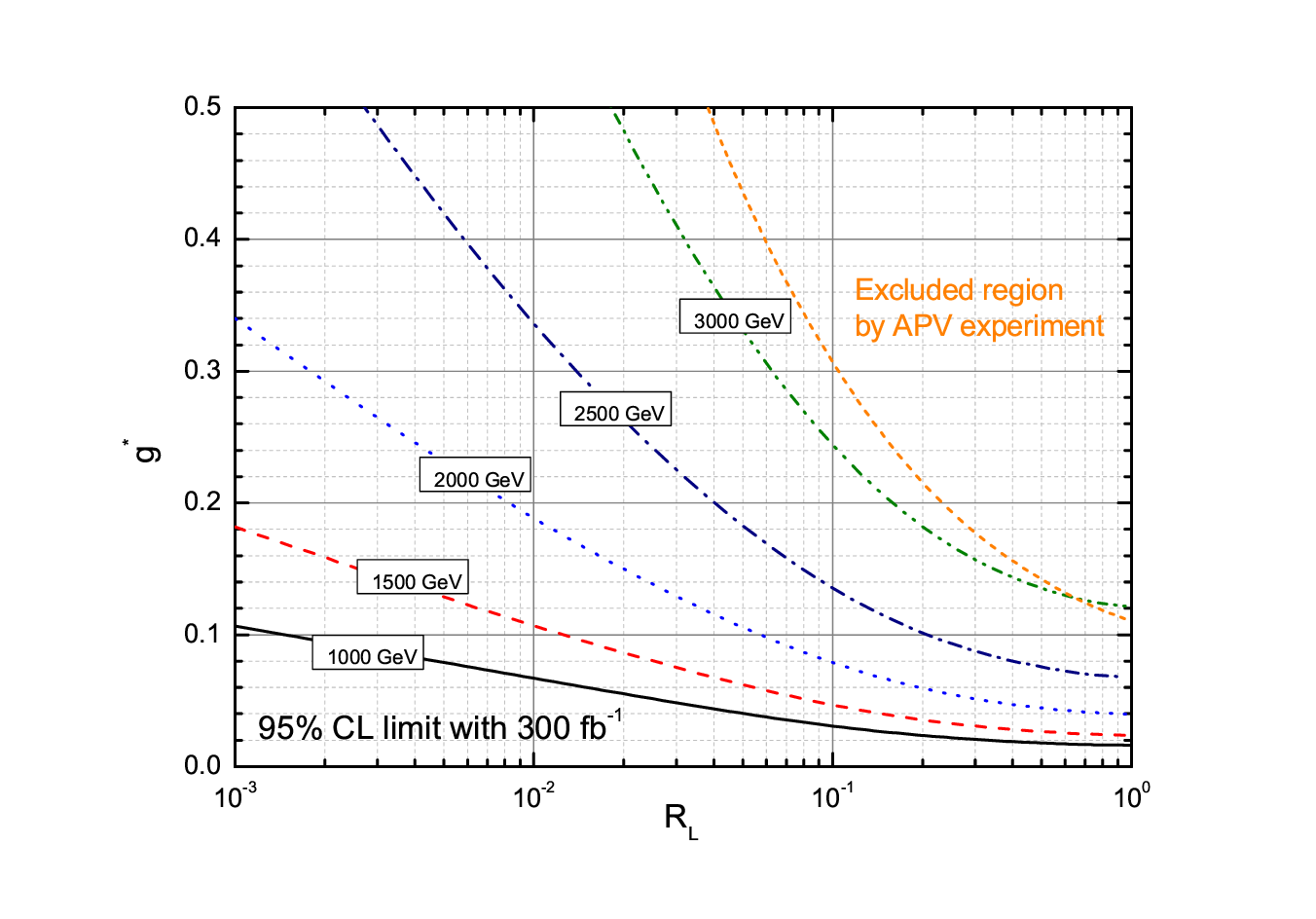}\epsfxsize=9cm \epsffile{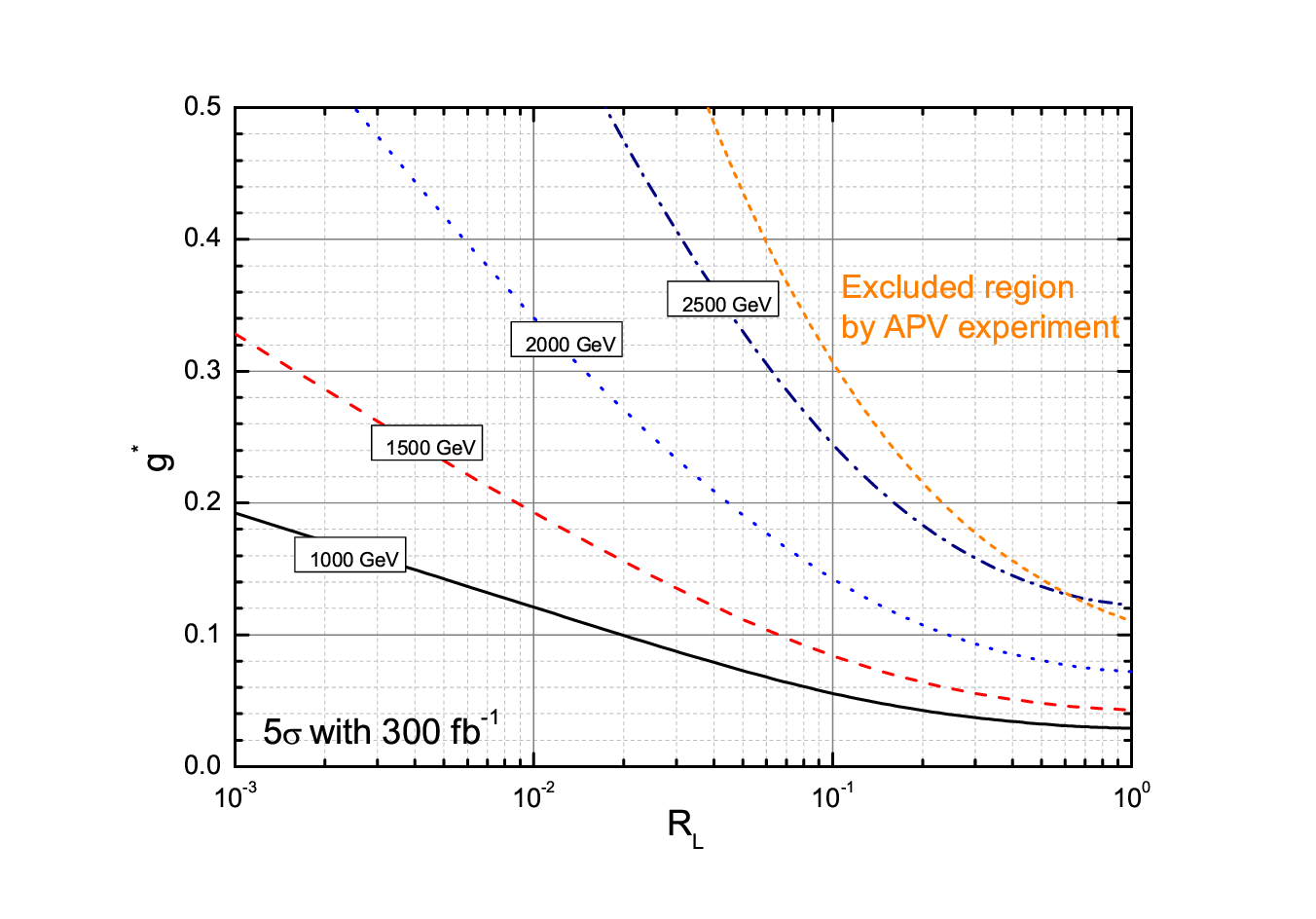}}
\centerline{\epsfxsize=9cm \epsffile{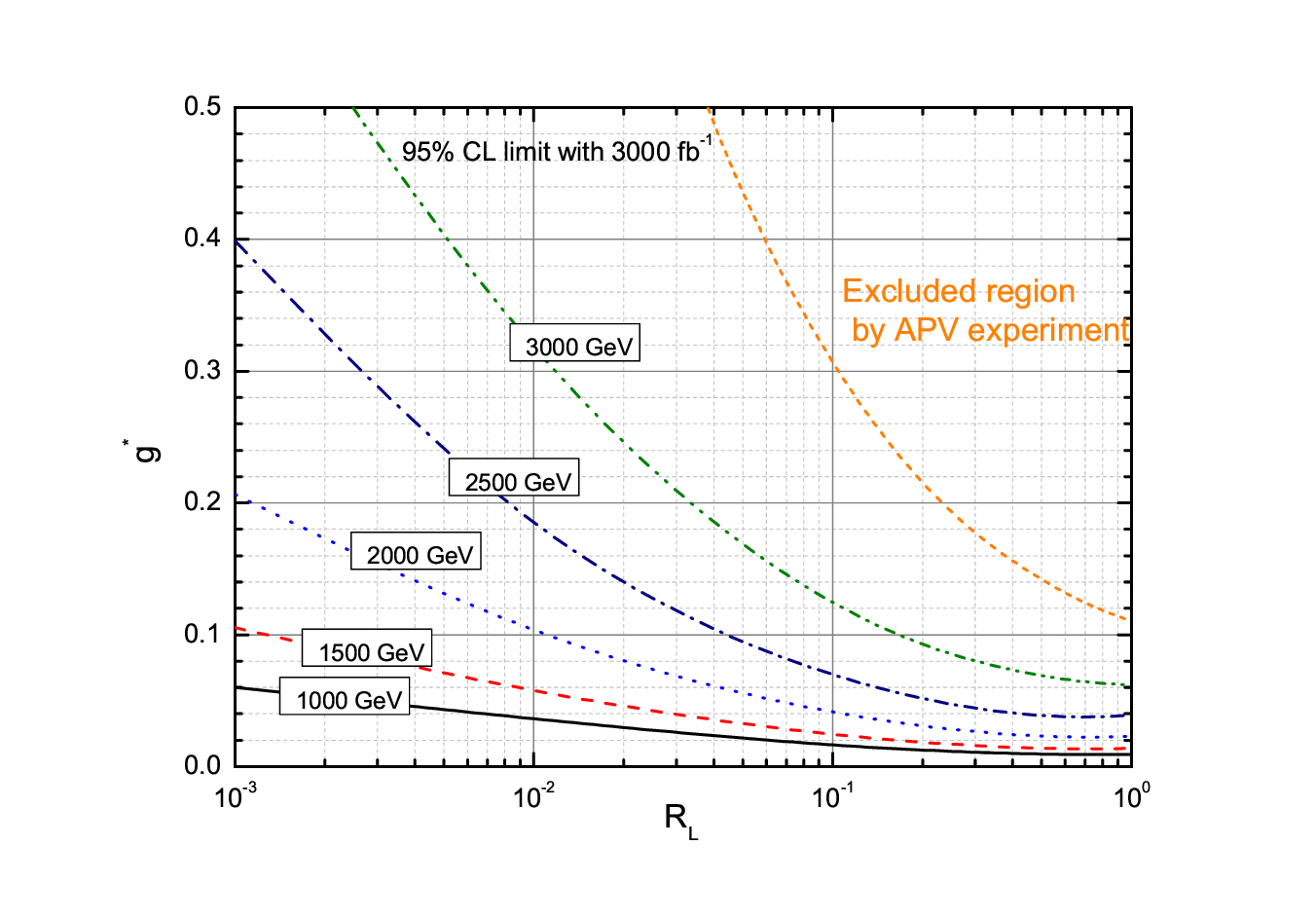}\epsfxsize=9cm \epsffile{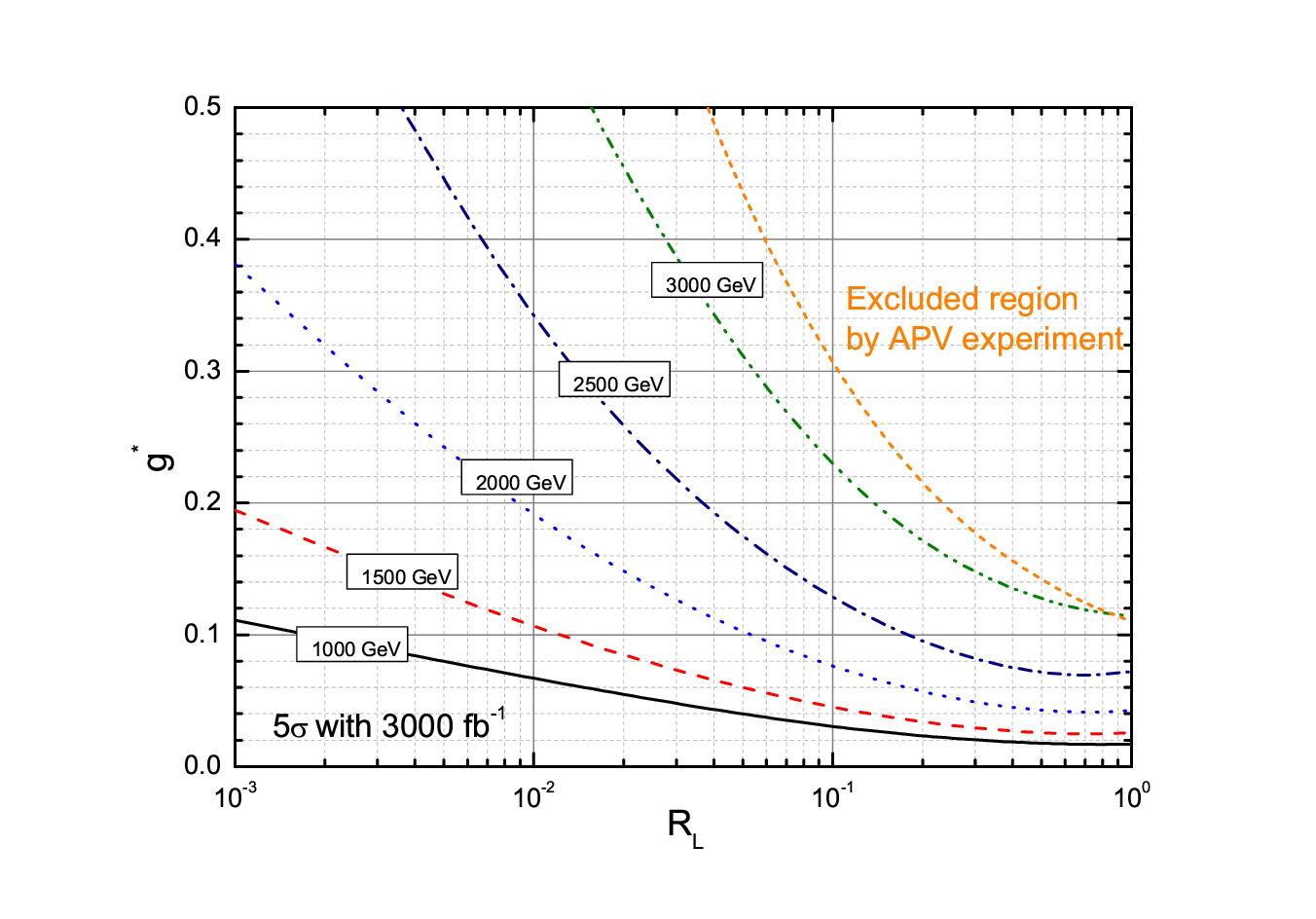}}
\caption{95\% CL exclusion limit (left panel) and $5\sigma$ (right panel) contour plots for the signal in $g^{*}-R_L$ for five typical mass parameters at 14 TeV LHC with an integrated luminosity of 300 fb$^{-1}$ (upper) and 3000 fb$^{-1}$ (down). Here we take a mild systematic of $\delta=10\%$. }
\label{ss-RL}
\end{center}
\end{figure}
The sensitivity that graphicized as contours in $g^{*}-R_{L}$ plane is presented in Fig.~\ref{ss-RL} with $\delta=10\%$ and five fixed typical VLQ-$X$ masses at the 14 TeV LHC with an integrated luminosity of 300 and 3000 fb$^{-1}$, respectively. The current limits from the  APV experiment are also displayed as dot-dashed curves. One can see that, for $M_X=1500, 2000, 2500$ GeV and $R_L=0.1$, the $5\sigma$ level discovery sensitivity of $g^{\ast}$ is respectively about $0.08, 0.14, 0.24$ with an integrated luminosity of $300 $ fb$^{-1}$, and changed as about $0.05, 0.14, 0.13$ with an integrated luminosity of  $3000$ fb$^{-1}$. Otherwise, the 95\% CL excluded region for the coupling parameter $g^{*}$ is
respectively about $0.05~(0.025), 0.08~(0.04), 0.14~(0.07)$ with an integrated luminosity of  $300~(3000)$ fb$^{-1}$.

\section{CONCLUSION}
We have present a study of the single production of  VLQ-$X$ at the future 14 TeV LHC. The work is performed in a simplified model that the SM extended with an SU(2) doublet \X assuming the VLQ-$X$ coupling only to the first- and preferentially
to third-generation quarks.
 We presented a search strategy at the future HL-LHC for a distinguishable signal with a same-sign dilepton plus one $b$-tagged jet and missing energy.  After performing a detector level simulation for the signal and relevant SM backgrounds, the $5\sigma$ discovery prospects and 95\% CL exclusion limits in the parameter plane were, respectively,  obtained at 14 TeV LHC with an integral luminosity of 300~(3000)~fb$^{-1}$, as displayed in Table~\ref{sum}. Assuming $Br(X\to tW)=1$, the authors in Refs.~\cite{Shang:2023ebe,Han:2023jzm} investigated the expected
limits for the VLQ-$X$ via the single production of $X$ at the LHC.
Therefore, we also present these results in Table~\ref{sum}, where the systematic uncertainty is taken as $\delta=30\%$ in Ref.~\cite{Shang:2023ebe} and $\delta=20\%$ in Ref.~\cite{Han:2023jzm}, respectively.

\begin{table}[htb]
\centering %
\caption{The 95\% CL exclusion limits and 5$\sigma$ signal discoveries at the HL-LHC. The systematic uncertainty is taken as $\delta=10\%$. }
\vspace{0.8cm}
\begin{tabular}{p{1.6cm}<{\centering} p{2.0cm}<{\centering} p{2.5cm}<{\centering} p{2.5cm}<{\centering}p{2.5cm}<{\centering} p{2.5cm}<{\centering} }
\toprule[1.5pt]
\hline
 \multirow{2}{*}{$R_L$}&\multirow{2}{*}{Luminosity}& \multicolumn{2}{c}{Exclusion}&\multicolumn{2}{c}{Discovery}  \\ \cline{3-6}
&(fb$^{-1}$)&$g^{*}$&$M_{X}(\rm GeV)$&$g^{*}$&$M_{X}(\rm GeV)$ \\      \cline{1-6}
 \multirow{2}{*}{0}&300&[0.12, 0.4]&[1000, 1950]&[0.21, 0.4]&[1000, 1550]\\ \cline{2-6}
 &3000&[0.06, 0.4]&[1000, 2450]&[0.1, 0.4]&[1000, 2000]\\ \hline
  \multirow{2}{*}{0.01}&300&[0.07,  0.4]&[1000, 2650]&[0.12, 0.4]&[1000, 2250]\\ \cline{2-6}
 &3000&[0.034, 0.29]&[1000, 3000]&[0.063, 0.4]&[1000, 2700]\\ \hline
  \multirow{2}{*}{0.1}&300&[0.027, 0.21]&[1000, 3000]&[0.05, 0.38]&[1000, 3000]\\ \cline{2-6}
 &3000&[0.014, 0.1]&[1000, 3000]&[0.026, 0.2]&[1000, 3000]\\ \hline
  \multirow{2}{*}{1}&300&[0.016, 0.12]&[1000, 3000]&[0.003, 0.22]&[1000, 3000]\\ \cline{2-6}
 &3000&[0.008, 0.06]&[1000, 3000]&[0.015, 0.11]&[1000, 3000]\\ \hline \hline
 \multirow{2}{*}{Ref.~\cite{Shang:2023ebe}}&300&[0.144, 0.2]&[1000, 1230]&$\setminus$&$\backslash$\\ \cline{2-6}
 &3000&[0.082, 0.2]&[1000, 1680]&[0.168, 0.2]&[1000, 1120]\\ \hline
 Ref.~\cite{Han:2023jzm}&3000&[0.12, 0.21]&[1300, 2000]&[0.23, 0.4]&[1300, 2000]\\
\hline
\end{tabular}\label{sum}
 \end{table}

Considering a systematic uncertainty of 10\%, the 14 TeV LHC with an integrated luminosity of 300~fb$^{-1}$  can discover  the correlated regions of $g^{*}\in [0.1, 0.4]~([0.015,0.22])$ and $M_{X}\in [1000, 1550]~[1000,3000]$~GeV for $R_{L}=0~(1)$.
On the other hand, the 95\% CL exclusion limits are $g^{*}\in [0.12, 0.4]~([0.01,0.12])$ and $M_{X}\in [1000, 1950]~[1000,3000]$~GeV for $R_{L}=0~(1)$. Meanwhile, the future HL-LHC with an integrated luminosity of 3000~fb$^{-1}$  can discover  the correlated regions of $g^{*}\in [0.1, 0.4]~([0.015,0.11])$ and $M_{X}\in [1000, 2000]~[1000,3000]$~GeV for $R_{L}=0~(1)$.
On the other hand, the 95\% CL exclusion limits are $g^{*}\in [0.06, 0.4]~([0.01,0.06])$ and $M_{X}\in [1000, 2450]~[1000,3000]$~GeV for $R_{L}=0~(1)$.
 We expect that our investigation will represent complementary
explorations for a potential VLQ-$X$ at  the HL-LHC.

\begin{acknowledgments}
This work is supported by the National Natural Science Foundation of China (Grant
No. 11904082).
\end{acknowledgments}


\end{document}